\documentclass[twocolumn]{aastex62}

\newcommand\teff{$T_{\mathrm{eff}}$}
\newcommand\logg{$\log g$}

\newcommand\teffeq{T_{\mathrm{eff}}}

\newcommand\rp{$R_{\mathrm{p}}$}
\newcommand\rpeq{R_{\mathrm{p}}}
\newcommand\rearth{$R_{\oplus}$}

\newcommand{\feh}{\mbox{$\rm{[Fe/H]}$}}
\newcommand{\kp}{$Kp$}
\newcommand{\kep}{\mbox{\textit{Kepler}}}
\newcommand{\gaia}{\mbox{\textit{Gaia}}}
\newcommand{\fearth}{$F_\oplus$}

\newcommand{\msun}{\ensuremath{\,M_\odot}}

\newcommand{\tess}{{\it TESS}}

\newcommand{\ktwo}{{K2}}

\newcommand{\nstars}{\mbox{177,911}}
\newcommand{\ndwarfs}{\mbox{120,000}}
\newcommand{\nsubg}{\mbox{37,000}}
\newcommand{\ngiants}{\mbox{21,000}}
\newcommand{\nbinary}{\mbox{3,100}}

\newcommand{\fdwarfs}{\mbox{67}}
\newcommand{\fsubg}{\mbox{21}}
\newcommand{\fgiants}{\mbox{12}}

\usepackage{amsmath}

\graphicspath{{./}{figures/}}

\shorttitle{Revised \kep\ Radii Using $Gaia$ DR2}
\shortauthors{Berger et al.}

\begin{document}

\title{Revised Radii of \kep\ Stars and Planets Using \gaia\ Data Release 2}

\correspondingauthor{Travis Berger}
\email{taberger@hawaii.edu}

\author[0000-0002-2580-3614]{Travis A. Berger}
\affiliation{Institute for Astronomy, University of Hawaii, 2680 Woodlawn Drive, Honolulu, Hawaii 96822, USA}

\author[0000-0001-8832-4488]{Daniel Huber}
\affiliation{Institute for Astronomy, University of Hawaii, 2680 Woodlawn Drive, Honolulu, Hawaii 96822, USA}
\affiliation{Sydney Institute for Astronomy (SIfA), School of Physics, University of Sydney, NSW 2006, Australia}
\affiliation{SETI Institute, 189 Bernardo Avenue, Mountain View, CA 94043, USA}
\affiliation{Stellar Astrophysics Centre, Department of Physics and Astronomy, Aarhus University, Ny Munkegade 120, DK-8000 Aarhus C, Denmark}

\author[0000-0002-5258-6846]{Eric Gaidos}
\affiliation{Department of Geology \& Geophysics, University of Hawaii at M\={a}noa, Honolulu, HI 96822, USA}

\author[0000-0002-4284-8638]{Jennifer L. van Saders}
\affiliation{Institute for Astronomy, University of Hawaii, 2680 Woodlawn Drive, Honolulu, Hawaii 96822, USA}

\begin{abstract}

\noindent
One bottleneck for the exploitation of data from the \kep\ mission for stellar astrophysics and exoplanet research has been the lack of precise radii and evolutionary states for most of the observed stars. We report revised radii of \nstars\ \kep\ stars derived by combining parallaxes from $Gaia$ Data Release 2 with the DR25 \kep\ Stellar Properties Catalog. The median radius precision is $\approx$\,8\%, a typical improvement by a factor of 4-5 over previous estimates for typical \kep\ stars. We find that $\approx$\,\fdwarfs\% ($\approx$\,\ndwarfs) of all \kep\ targets are main-sequence stars, $\approx$\,\fsubg\% ($\approx$\,\nsubg) are subgiants, and $\approx$\,\fgiants\% ($\approx$\,\ngiants) are red giants, demonstrating that subgiant contamination is less severe than some previous estimates and that \kep\ targets are mostly main-sequence stars. Using the revised stellar radii, we recalculate the radii for 2123 confirmed and 1922 candidate exoplanets. We confirm the presence of a gap in the radius distribution of small, close-in planets, but find that the gap is mostly limited to incident fluxes $>$\,200\,\fearth\ and its location may be at a slightly larger radius (closer to $\approx$\,2\,\rearth) when compared to previous results. Further, we find several confirmed exoplanets occupying a previously-described ``hot super-Earth desert'' at high irradiance, show the relation between gas-giant planet radius and incident flux, and establish a bona-fide sample of eight confirmed planets and 30 planet candidates with \rp\,$<$\,2\,\rearth\ in circumstellar ``habitable zones" (incident fluxes between 0.25--1.50\,$F_\oplus$). The results presented here demonstrate the potential for transformative characterization of stellar and exoplanet populations using \gaia\ data.

\end{abstract}


\section{Introduction} \label{sec:intro}

Precise estimates of exoplanet properties such as radius, mass, and density inevitably require precise characterization of the host stars.  Precise stellar classifications are also required to  study the dynamics and evolution of planetary orbits  \citep{kane12,sliski14,vaneylen15,Shabram2016} and derive an accurate planet occurrence \citep[e.g.][]{Howard2012,burke15}.

Traditional methods used to classify the target stars of exoplanet surveys include broadband colors and proper motions, which efficiently separate dwarfs from giants but cannot resolve intermediate evolutionary states, with typical uncertainties of $\approx$\,0.3--0.4\,dex in \logg\ \citep{brown11,huber16}. High-resolution spectroscopy delivers typical precisions of  $\approx$\,0.15\,dex in \logg\ \citep{torres12} for solar-type stars, while methods calibrated to benchmark stars can achieve precisions down to $\approx$\,0.07\,dex \citep{brewer15,petigura15b}. Finally, time-domain variability of stars offers currently the highest precision \logg\ values for field stars, for example by measuring amplitudes or timescales of stellar granulation ($\approx$\,0.1\,dex, \citeauthor{bastien13} \citeyear{bastien13}; $\approx$\,0.03\,dex, \citeauthor{kallinger16} \citeyear{kallinger16}) or stellar oscillations \citep[$\approx$\,0.01\,dex,][]{huber13}. 

Despite this progress, most of these methods are only applicable to a subset of the large samples of stars that are typically observed in exoplanet transit surveys (190,000 stars for \kep, $>$\,200,000 stars for \ktwo, $>$500,000 stars for the Transiting Exoplanet Survey Satellite (\tess)). As a result, 70\% of the overall \kep\ population in the latest version of the \kep\ Stellar Properties Catalog \citep[KSPC DR25,][]{Mathur2017} still have \logg\ values determined from photometry. This translates into 30--40\% uncertainties in stellar radii that are severely limiting our understanding of the stellar and planet population probed by \kep.

Improved stellar radii of \kep\ hosts have recently led to several important results for our understanding of exoplanets, such as the discovery of a gap in the distribution of small planets by the California-\kep\ Survey \citep[CKS,][]{Fulton2017, Petigura2017,Johnson2017} and evidence for a dearth of hot super-Earths \citep{Lundkvist2016}. Both results have been tied to processes such as photoevaporation \citep{Lopez2012,Owen2017}, but are limited subsamples consisting of less than half of planet candidates.

The bottleneck caused by imprecise stellar radii of \kep\ stars can now be relieved thanks to precise parallaxes from \gaia\ Data Release 2 (DR2) for over one billion stars in the galaxy \citep{Brown2018,Lindegren2018}. In this paper we re-derive radii for \nstars\ \kep\ stars using \gaia\ DR2 parallaxes, and investigate the stellar and exoplanet radius distributions of \kep\ targets.

\section{Methodology} \label{sec:methods}

\subsection{\kep-\gaia\ DR2 Cross-matching} \label{sec:cross}

First, we cross-matched the positions of all stars from the KSPC DR25 \citep{Mathur2017} by utilizing the X-match service of the Centre de Donn\'{e}es astronomiques de Strasbourg (CDS). This provided a table of \gaia\ DR2 source matches within three arcseconds of each \kep\ star. To determine bona-fide \kep-$Gaia$ source matches, we first removed all matches with distances greater than 1.5 arcseconds from the \kep-determined position. We chose 1.5 arcseconds because the distribution of separations displayed a minimum there, and the increase of matches at greater angular separations indicates the inclusion of spurious background sources. 

Next, we imposed a variety of magnitude cuts, depending on the available photometry, to ensure our \kep-\gaia\ matches were of similar brightness. Unfortunately, not all \kep\ stars had similar quality photometry to compare to the measured \gaia\ $G$-band magnitudes, so we had to utilize AAVSO Photometric All-Sky Survey (APASS) $g$, $r$, and/or $i$ photometry for instances where KSPC stars did not have $g$-, $r$-, or $i$- band photometry from the \kep\ Input Catalog \citep[KIC,][]{brown11}. For stars that were still missing any $g$, $r$, or $i$ photometry, we used \kep\ magnitudes (\kp) for comparisons with $G$ magnitudes.

To compute our predicted $G$ magnitudes, we utilized the $g$, $r$, and $i$ color-color polynomial fits in Table 7 of \cite{jordi10}. After inspecting the distribution of $G_{Gaia}$--$G_{\mathrm{pred}}$, we chose to remove all stars with absolute differences greater than two magnitudes. For the remaining sample of stars with only \kp\ magnitudes, we compared $G_{Gaia}$--\kp\ and again removed all stars with absolute differences greater than two magnitudes.

For stars with multiple matches that satisfied these criteria, we decided to keep those with the smallest angular separations. Of the 197,104 stars present in the KSPC, we identified \gaia\ DR2 source matches for 195,710. Stars with poorly determined parallaxes ($\sigma_{\pi}/\pi$\,$>$\,0.2), low effective temperatures based on our adopted values (\teff\,$<$\,3000\,K, see Section 2.2), extremely low \logg\ ($<$\,0.1\,dex), and/or non-``AAA''-quality Two Micron All Sky Survey (2MASS) photometry were rejected from our sample.

Additionally, we made astrometric cuts similar to those described in Appendix C of \cite{Lindegren2018} and Section 4.1 of \cite{Arenou2018}. In particular, we used Equation (1) (unit weight error compared to a function of the $G$ magnitude of the source that helps filter contamination from binaries and calibration problems) and Equation (3) (greater than eight groups of observations separated by at least 4 days) of \cite{Arenou2018} to remove stars with bad astrometric solutions. We did not use the astrometric excess noise values provided by \gaia\ DR2 to filter stars because they were less discriminating for stars with $G$\,$<$\,15 due to the ``degree of freedom bug" \citep[see Appendix A and C of][]{Lindegren2018}. We did not use Equation (2) of \cite{Arenou2018}, a cut ensuring that $Gaia$ has clean photometry of the included sources, because we utilized separate 2MASS photometry in our analysis. As discussed in \cite{Lindegren2018}, our imposed cuts removed many stars that appear in unphysical areas of radius-\teff\ parameter space, such as the ``subdwarfs'' between the stellar main sequence and the white dwarf branch. Excluding these stars reduced our final sample to \nstars\ \kep\ stars.

\subsection{Stellar Radii Determination} \label{sec:raddet}

\begin{figure}
\resizebox{\hsize}{!}{\includegraphics{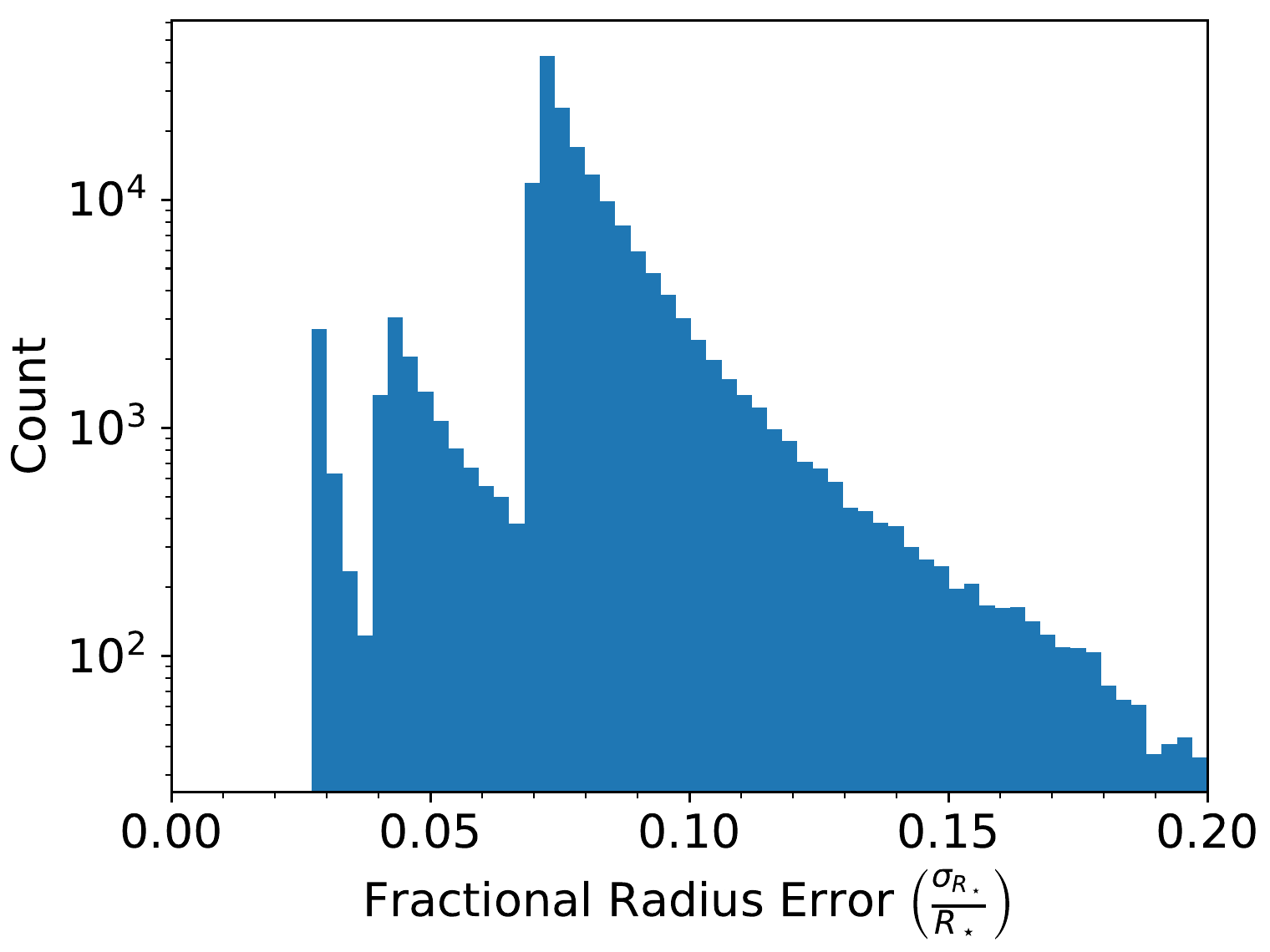}}
\caption{Histogram of the fractional radius uncertainty for 177,702 \kep\ stars derived in this work. The sample of 209 stars with fractional radius uncertainties $>0.2$ are some of the most distant stars in the \kep\ field. The typical radius uncertainty pre-$Gaia$ DR2 was $\approx$\,30\%. The peaks at $\approx$\,3\%, $\approx$\,4.5\%, and $\approx$\,8\% errors correspond to M-dwarfs with radii determined from $M_{K_s}$--radius relations, stars with spectroscopic constraints on \teff, and stars with photometric \teff, respectively.} 
\label{fig:rade}
\end{figure}

To calculate stellar radii we employed the stellar classification code \texttt{isoclassify} \citep{huber17} in its ``direct method,'' using as input the \gaia\ DR2 parallax \citep{Lindegren2018}, 2MASS K-band magnitude, and \teff, \logg, \feh\ values from the DR25 KSPC \citep{Mathur2017}. We replaced the input values given in the KSPC for two sets of stars: stars in the California-\kep\ Survey (CKS), for which we adopted spectroscopic parameters from \citet{Petigura2017}, and stars with \teff\,$<$\,4000\,K with \teff\ provenances from the KIC, for which we adopted revised \teff\ values from \citet{Gaidos2016}.

For each star, we first converted parallaxes into distances using an exponentially decreasing volume density prior with a length scale of 1.35\,kpc \citep{bailer15,astraatmadja16} and included a systematic parallax offset of 0.03\,mas \citep{Lindegren2018}. We note that $Gaia$ DR2 has systematic parallax offsets that vary with position, angular scale, and color \citep{Arenou2018,Lindegren2018}. \cite{Zinn2018} used asteroseismology to compare distances to those derived from $Gaia$ parallaxes, and found a systematic offset of 0.05\,mas within the $Kepler$ field. Although this measurement applies to the $Kepler$ field, we still used the \cite{Lindegren2018} value of 0.03\,mas derived from quasars because of potential systematics in asteroseismic scaling relations and poorly constrained color dependencies in the parallax offset. In addition, the 0.02\,mas offset was small compared to the median parallax of 0.66\,mas in our sample.

We then combined the 2MASS $K$-band magnitude with extinctions $A_{V}$ derived from the 3D reddening map and interpolated reddening vectors in Table 1 of \cite{Green2018}. We also added the gray component of the extinction curve $b$\,=\,0.063, computed from $A_H$/$A_K$\,=\,1.74 \citep{Nishiyama2006} by \cite{Green2018}, to our extinction values. Next, we added these extinction values to our magnitudes, which we then combined with distances to calculate absolute magnitudes. We derived bolometric corrections by linearly interpolating \teff, \logg, \feh\ and $A_{V}$ in the bolometric correction tables from the MESA Isochrones \& Stellar Tracks \citep[MIST,][]{choi16} grids (MIST/C3K, Conroy et al., in prep\footnote{\url{http://waps.cfa.harvard.edu/MIST/model_grids.html}}), which we combined with our absolute magnitudes to compute luminosities. Finally, we combined the derived luminosities with \teff\ from \citet{Mathur2017} (or other sources as indicated above), and \gaia\ parallaxes in the Stefan-Boltzmann relation to calculate stellar radii. The procedure is implemented as a Monte-Carlo sampling scheme, and the resulting distributions were used to calculate the median and 1\,$\sigma$ confidence interval for the radius of each star. Table \ref{tab:stars} lists our revised radii for all \nstars\ \kep\ stars analyzed here.

The above method produced systematically overestimated radii for M-dwarfs due to inaccuracies in bolometric corrections in \texttt{isoclassify}, which are based on ATLAS model stellar atmospheres \citep{Kurucz1993}. Therefore, we used an empirical relationship between the absolute $K$ magnitude ($M_{K_s}$) and stellar radius described by Equation (4) and Table 1 of \cite{mann15} to compute stellar radii and hence luminosities for stars with \teff\,$<$\,4100\,K and $M_{K_s}$\,$>$\,3.0\,mag. We added 2.7\%, corresponding to the uncertainty of the relation, to uncertainties in the radii of these stars. Although the $M_{K_s}$--radius relation only applies for $M_{K_s}$\,$>$\,4.0\,mag, we have confirmed that an extrapolation to $M_{K_s}$\,=\,3.0\,mag produces radii that are approximately compatible with those predicted by MIST isochrones.

Figure \ref{fig:rade} shows a histogram of fractional radius uncertainties for 177,702 of \nstars\ \kep\ stars with radii derived in this work. The remaining 209 stars have higher fractional radius uncertainties, and are likely some of the most distant stars in the \kep\ field. The typical uncertainty is $\approx$\,8\%, a factor of 4-5 improvement over the KSPC. The radius uncertainty is dominated by \teff, which for a typical \kep\ target is $\approx$\,3.5\% based on broadband photometry \citep{huber14}. The peak at $\approx$\,3\% fractional radius uncertainty corresponds to M-dwarf radii computed through the $M_{K_s}$--radius relation \citep[not dependent on \teff,][]{mann15}, while the peak at $\approx$\,4.5\% fractional radius uncertainty represents stars with spectroscopic temperatures (2\% errors in \teff). Our error budget also included uncertainties of 0.04\,mag in $A_{V}$ and 0.02\,mag in bolometric corrections, which are typical values for the \kep\ field \citep{huber17}. To compute the uncertainty in $A_V$, which carries into the error in the stellar radius, we combined both the distance uncertainty, which translates into an uncertainty in $A_V$ along the line of sight (minimal), and uncertainties in the reddening model itself (dominant). We determined the latter by comparing the \cite{green15} map with the \cite{Green2018} map for our sample, yielding a median absolute deviation of $\sim$\,22\%, which we adopt as a fractional uncertainty for our reported extinction values. Therefore, for our typical $A_V$\,=\,0.18\,mag, we report a typical uncertainty of 0.04\,mag. This corresponds to $A_K$\,=\,0.013\,$\pm$\,0.003\,mag, which factors into the absolute $K$-band magnitude uncertainty and hence our stellar radius uncertainty. We emphasize that the above routine uses \logg\ from the KSPC only to determine bolometric corrections, which are only mildly dependent on \logg\ and hence the derived radii are mostly insensitive to inaccurate \logg\ values.

The 3.5\% and 2\% uncertainties in \teff\ ($\approx$\,200\,K and $\approx$\,115\,K at solar \teff) were conservative, but large enough to have encompassed systematic differences between \teff\ scales and covariances between extinction and color-\teff\ relations \citep{pinsonneault11}. Future revisions of the \teff\ scale for \kep\ stars, taking into account revised reddening maps based on $Gaia$ DR2, can be expected to improve the typical radius precision to $\sim$\,5\% or better.

The $Gaia$ Collaboration released radii and effective temperatures for 178,706  \kep\ targets based on \gaia\ photometry and parallaxes \citep{Brown2018,Lindegren2018,Andrae2018}. However, these parameters are optimized for $>$\,160 million stars across the sky. In contrast, the \kep\ field is one of the most well-studied samples of stars due to its relevance to exoplanet science, and the KSPC includes information from the vast amount of photometric, spectroscopic and asteroseismic analyses that have been performed over the past ten years. Therefore, we expect the stellar radii derived in this work to be more accurate than those reported by the $Gaia$ Collaboration.

\subsection{Validation of Stellar Radii} \label{sec:valid}

\subsubsection{Comparison to Asteroseismic Radii} \label{sec:astcomp}

\begin{figure}
\resizebox{\hsize}{!}{\includegraphics{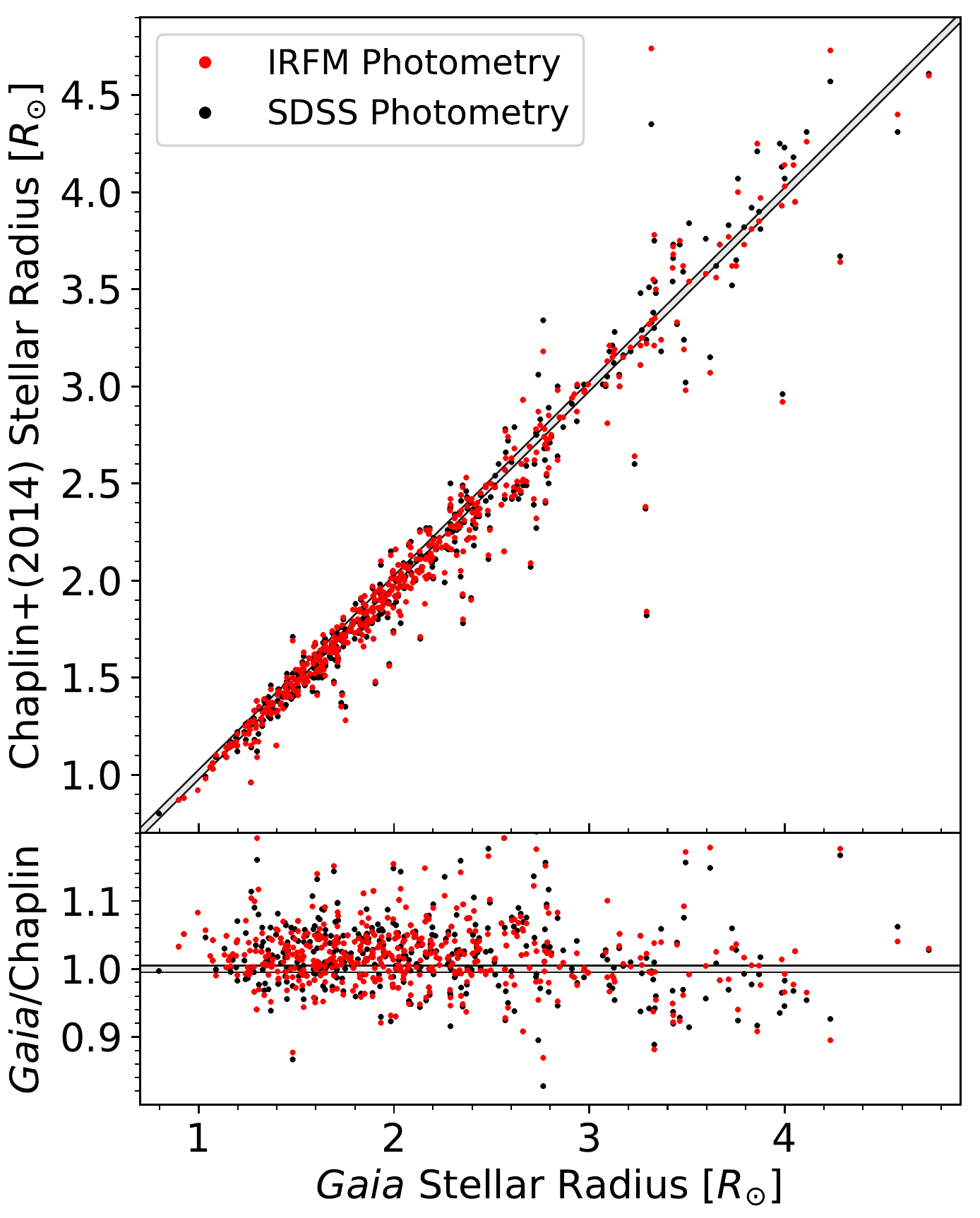}}
\caption{Comparison of our derived stellar radii using Gaia parallaxes to asteroseismic radii from \cite{Chaplin2014}. In red and black are \cite{Chaplin2014} radii derived from \teff\ determined through the InfraRed Flux Method (IRFM) and SDSS photometry, respectively. The top panel plots \cite{Chaplin2014} radii versus those derived in this work, while the bottom panel plots the ratio (our radii divided by the Chaplin radii) versus our radii.} 
\label{fig:chapcomp}
\end{figure}

To test the precision of our radii, we compared them to radii derived using asteroseismology as given in \cite{Chaplin2014} (Figure \ref{fig:chapcomp}). Red and black points represent \cite{Chaplin2014} radii determined from \teff\ derived from the InfraRed Flux Method (IRFM) and SDSS photometry, respectively. Temperatures adopted in our catalog come from spectroscopic measurements by \citet{buchhave15}, as adopted by \citet{Mathur2017}. Overall we find that the scatter is on the order of $\approx$\,4\%, which is fully consistent with the typical $\approx$\,4\% uncertainties of our radii for stars with spectroscopic constraints (see Figure \ref{fig:rade}). We also identify a $\approx$\,3\% systematic offset in the subgiant range (1.5--3.0\,$R_\odot$), where the \cite{Chaplin2014} radii are systematically smaller. Part of this offset can be explained by the use of different effective temperature scales, as discussed in \cite{huber17}, which identifies a similar offset based on a comparison of asteroseismology with $Gaia$ DR1. Ultimately, this comparison with independent measurements supports the precision of the radii reported in our catalog.

\subsubsection{Systematic Error Sources} \label{sec:acc}

A variety of factors can affect the accuracy of our reported stellar radii. Offsets in the effective temperature, in most cases, have the largest effect on our reported radii ($>$\,60\% of the error budget for a typical star with either spectroscopic, 2\%, or photometric, 3.5\% fractional errors in \teff). We used conservative errors on our \teff\ values because of the inhomogeneity of the KSPC's \teff\ sources. These uncertainties contained \teff\ offsets between different methods, which are typically less than 150\,K \citep[see Table 7 and \teff\ comparison plots in][]{Petigura2017}.

2MASS reports typical errors of 0.03\,mag in $K$-band photometry. Therefore, any systematic offset in the zeropoint of 2MASS photometry would, at most, result in a $\approx$\,1.5\% error in our computed stellar radius. \gaia\ DR2 parallaxes in the \kep\ field may be systematically underestimated by about 0.02\,mas, a figure smaller than typical formal error, as well as the global systematic value of 0.03\,mas \citep{Lindegren2018,Zinn2018}.  This offset would produce an overestimation of stellar radii of $\approx$\,1\% for nearby stars, and up to $\approx$\,5\% for stars as far as 5\,kpc. We included a 0.02\,mag uncertainty ($\lesssim$\,1\% error in the stellar radius) to account for uncertainty in our MIST/C3K bolometric correction grid, but that does not account for issues in the models. Although the grid appears to work well for most stars, it fails for M-dwarfs, where, in some cases, radii were overestimated by $\sim$\,20\%. We therefore computed M-dwarf radii using the \cite{mann15} relation detailed in Section \ref{sec:raddet}. Finally, we considered systematic errors in our extinction values. As we discussed in Section \ref{sec:raddet}, the bulk of the uncertainty in the extinction will come from intrinsic inaccuracies in the reddening map. Taking our typical extinction uncertainty of 0.003\,mag in $A_K$, this translates into a $<<$\,1\% underestimation of the stellar radius. Even when we considered the worst-case scenario from \cite{Green2018}, where their map significantly underestimates reddening by 0.25\,mag in $A_V$ ($A_K$\,=\,0.02\,mag) compared to the \citep{Planck2014} map, this only corresponds to a $\approx$\,1\% underestimation of the stellar radius.

In summary, we expect that individual systematic errors are well within our quoted uncertainties. Since some of the error sources are independent (e.g. temperature and parallax, photometric zero-point offsets and bolometric corrections) we consider it unlikely that they would be linearly additive, in which case radius systematics would exceed our quoted uncertainties.

\begin{figure*}
\resizebox{\hsize}{!}{\includegraphics{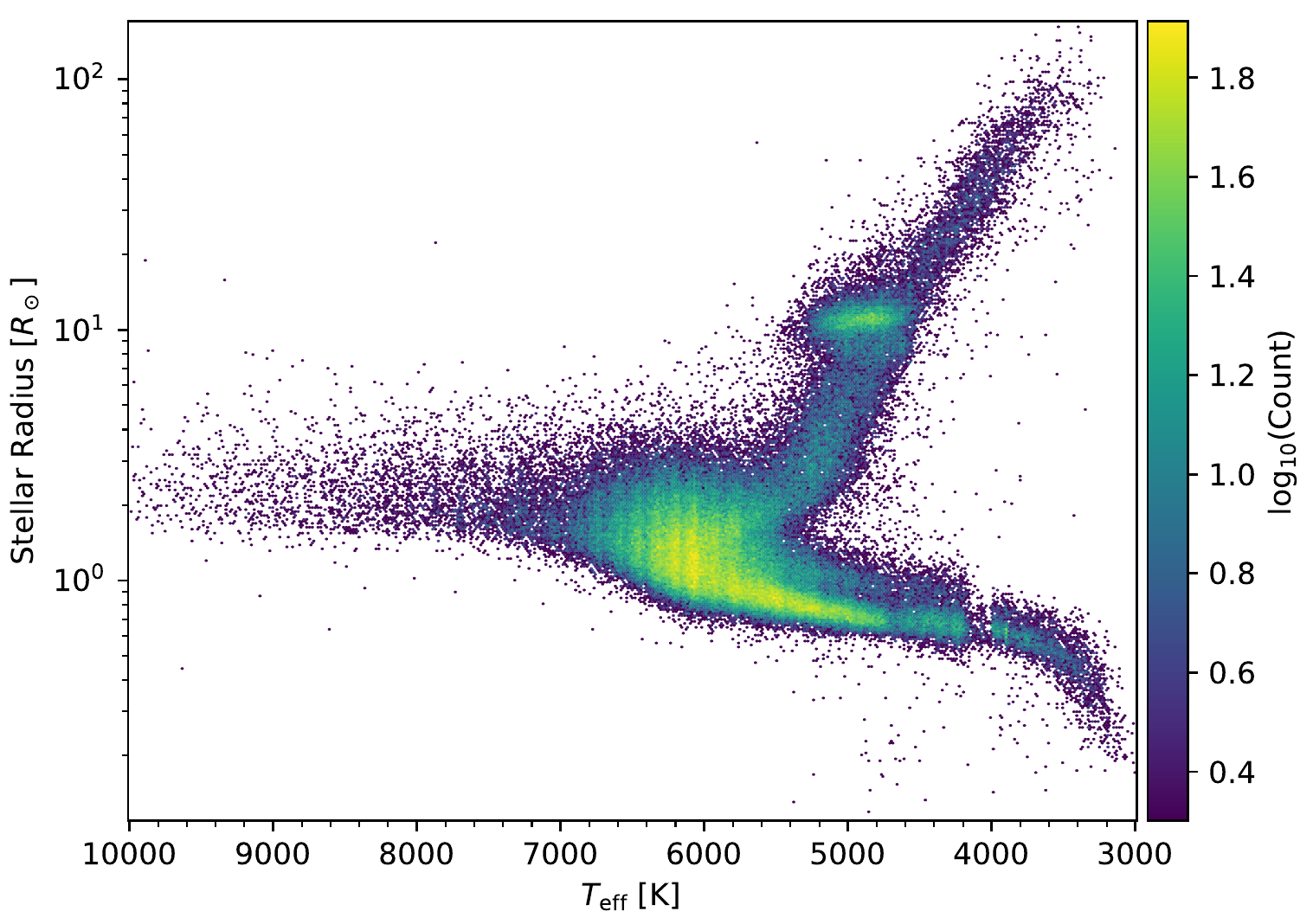}}
\caption{Radius versus effective temperature for 177722 \kep\ stars with radii based on $Gaia$ DR2 parallaxes presented in this work. A sample of 189 stars falling off the plot limits shown here includes hot stars (\teff\,$>$\,10000\,K) and white dwarfs. Color-coding represents logarithmic number density. Note that the discontinuity in \teff\ near 4000\,K is an artifact due to systematic shifts in \teff\ scales in the DR25 \kep\ Stellar Properties Catalog.} 
\label{fig:HR}
\end{figure*}

\subsubsection{Stellar and Exoplanet Radius Dilution} \label{sec:dilution}

2MASS photometry in some cases includes flux from unresolved stellar companions, which affects both stellar and exoplanet radii. To minimize the number of stars with problematic 2MASS photometry, we only used sources with ``AAA'' photometry quality, which removed 5,000 ($\approx$\,3\%) sources from our catalog. 2MASS photometry has an effective resolution of 4'' \citep{skrutskie06}, similar to the size of \kep\ pixels. \cite{Ziegler2018} showed that, of the companions within 4'' from their hosts, the contrasts ($\Delta$$m$) range over 0--6\,mag in the LP600 bandpass \citep[a long-pass filter that begins to transmit at 600\,nm and that roughly matches the \kep\ passband,][]{law14}. This corresponds to $\Delta$$m$\,$\approx$\,0--3\,mag for 2MASS $K$-band photometry for a G-type main sequence star and its companion, which results in a $\approx$\,41--3\% overestimated stellar radius for the primary star. This is significantly larger than our radius uncertainties in some cases, but, lacking adaptive optics follow-up for all stars in the \kep\ field, we did not amend our radii. \cite{Ziegler2018} found that $\approx$\,14.5\% of \kep\ stars with candidate planets have close-in ($<$\,4'') detected companions. However, only $\approx$\,7\% of stars in the \cite{Ziegler2018} sample had $\Delta$$K$\,$<$\,2\%. Thus, only these low-contrast companions could dilute measured fluxes enough to exceed our reported 8\% uncertainties. Companions more widely separated than 4'' should be resolved by 2MASS and in these cases the amount of dilution and affect on planet radius should be small.

If the stellar radius is actually smaller, then any transiting planet radius is smaller too. However, unresolved companions also affect the transit signal in the \kep\ lightcurve, and there is a net effect only to the extent the \emph{surface brightnesses} of the stars are different. For the \kep\ bandpass differences in \teff\ between the primary and companion will dominate, while differences in \logg\ and [Fe/H] will have minimal effect. We flagged stars identified as multiples by \citet{Ziegler2018} as adaptive optics (AO) binaries in Table \ref{tab:stars} (binary flag = 1 or 3). We caution that these flags are not complete as there may be companions unresolved by Robo-AO, they are restricted to the \kep\ Objects of Interest (KOIs), and not all detections are physical companions.

\section{Revised Radii of \kep\ stars}

\subsection{The $Gaia$ H-R Diagram of \kep\ Stars} \label{sec:comp}

Figure \ref{fig:HR} shows stellar radius versus effective temperature for the \kep\ stars with radii revised by this work. This diagram is the first nearly model-independent H-R diagram of the \kep\ field. We see a clear main sequence, from M dwarfs at \teff\,=\,3000\,K and $R$\,$\approx$\,0.2\,$R_\odot$, through A stars at \teff\,$\lesssim$\,9000\,K and $R$\,$\approx$\,2\,$R_\odot$. The main sequence turnoff at \teff\,$\approx$\,6000\,K and $R$\,$\approx$\,2\,$R_\odot$ is visible, along with the giant branch. We identify the ``red clump'' as the concentration of stars surrounding \teff\,$\approx$\,4900\,K and $R$\,$\approx$\,11\,$R_\odot$. As expected, the \kep\ targets are heavily dominated by FG-type stars as a result of the target selection focusing on solar-type stars to detect transiting exoplanets \citep{batalha10}.

The distribution in Figure \ref{fig:HR} contains artifacts, most prominently the gap in the main sequence around 4000\,K. This gap is the result of the combination of two  photometric \teff\ scales in the KSPC \citep{Mathur2017}, namely \teff\ values from \citet{pinsonneault11} for FGK stars and the classification of M dwarfs by \citet{dressing13}. An accurate re-calibration of the \teff\ scale for all \kep\ targets is beyond the scope of this paper, but the use of the DR25 ensures the inclusion of the best available values for \teff\ and \feh\ on a star-by-star basis.  A number of stars below the main sequence that may be white dwarfs (\teff\,=\,6500--10000\,K and $R$\,=\,0.02\,$R_\odot$, not shown in Figure \ref{fig:HR}) and subdwarfs (\teff\,=\,3600--5400\,K and $R_\star$\,$<$\,0.6\,$R_\odot$) as well as in other extreme parameter regimes could be catalog mismatches or have erroneous \teff\ values (Table \ref{tab:stars}).

Figure \ref{fig:HR} contains an apparent second sequence above the main sequence for dwarfs with \teff\,$<$\,5200\,K. Because K stars are less massive than their hotter main sequence counterparts, we do not expect these stars to have evolved significantly over a Hubble time, and the intrinsic spread in metallicity is not expected to be asymmetric enough to produce this feature. Rather, this feature likely contains nearly equal-mass binaries; the luminosities and radii of these stars will be overestimated by our methods, but also indicates that \gaia\ DR2 parallaxes can be used to readily identify cool main-sequence binaries.

\subsection{Comparison to the DR25 \kep\ Stellar Properties Catalog}

\begin{figure}
\resizebox{\hsize}{!}{\includegraphics{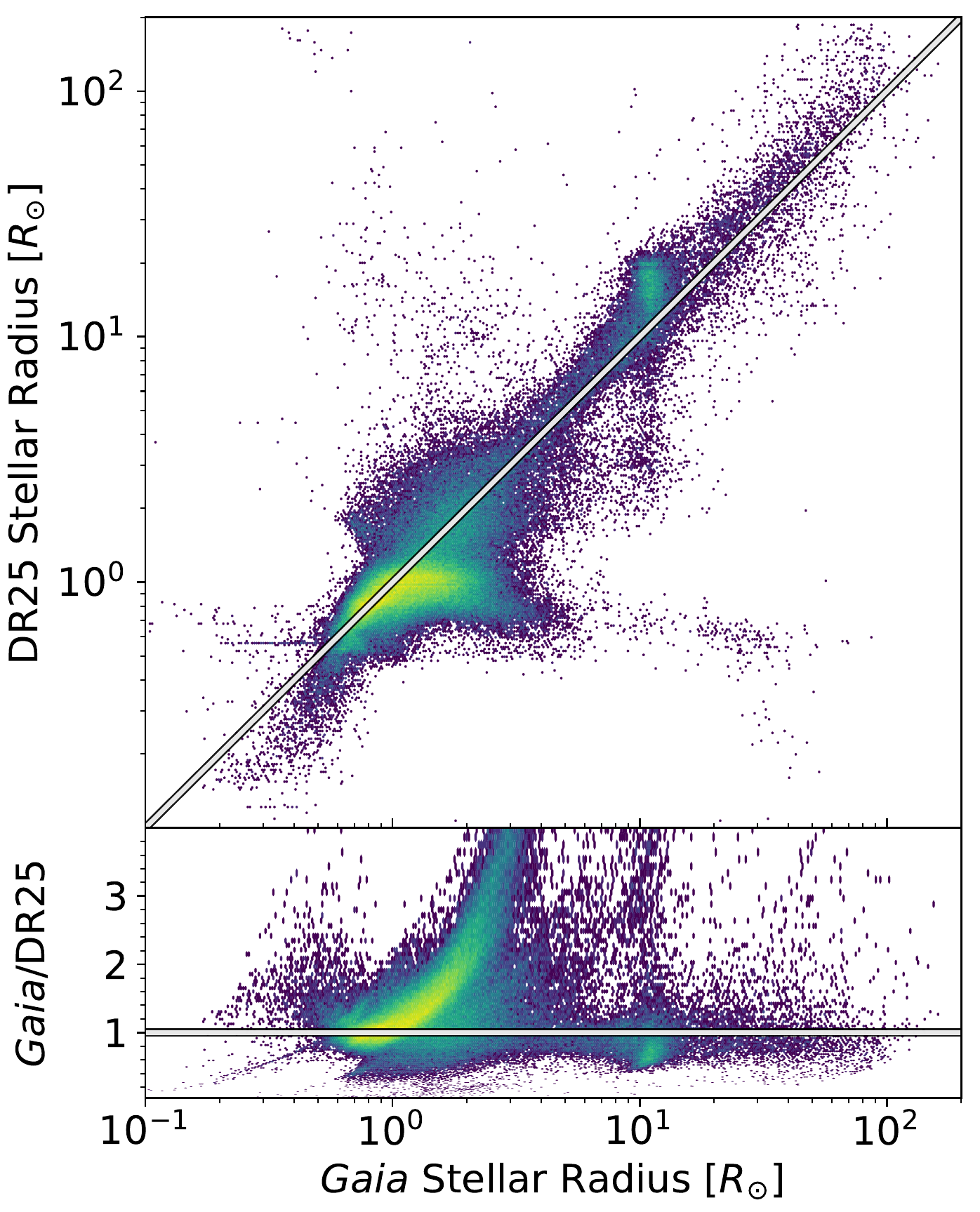}}
\caption{Comparison of radii in the DR25 \kep\ Stellar Properties Catalog \citep{Mathur2017} and the radii derived in this paper. The colors represent the density of points. The white and black line is the 1:1 comparison between DR25 radii and our derived radii. The bottom panel shows the ratio between DR25 stellar radii and our stellar radii.} \label{fig:radcomp}
\end{figure}

Figure \ref{fig:radcomp} shows a comparison of stellar radii in the DR25 stellar properties catalog \citep{Mathur2017} to those derived in this paper. The distribution approximately tracks the 1:1 line, but there is large scatter and strong systematic offsets caused by large uncertainties in the DR25 radii, which were mostly based on photometric \logg\ values from the KIC. We measure an overall median offset and scatter in the $Gaia$/DR25 residuals of 12\% and 34\% for all stars, 14\% and 32\% for unevolved stars ($<$\,3\,$R_\odot$), and --7\% and 35\% for red giants ($>$\,3\,$R_\odot$), where positive offsets indicate a larger \gaia\ radius. The residuals clearly demonstrate that a substantial fraction of \kep\ stars are more evolved than implied in the KSPC.

We also identify 975 giants which were misclassified as dwarfs and 483 dwarfs which were misclassified as giants (bottom right and top left areas in the top panel of Figure \ref{fig:radcomp}, respectively). The revised classifications presented here will thus aid in increasing cool dwarf samples for studies of stellar rotation and activity \citep[e.g.][]{McQuillan2014,angus15,davenport16} and red giants for asteroseismology \citep[e.g.][]{hekker11, mosser11c, stello13, yu18}.

\subsection{Evolutionary States of \kep\ Stars} 

\begin{figure}
\centering
\resizebox{\hsize}{!}{\includegraphics{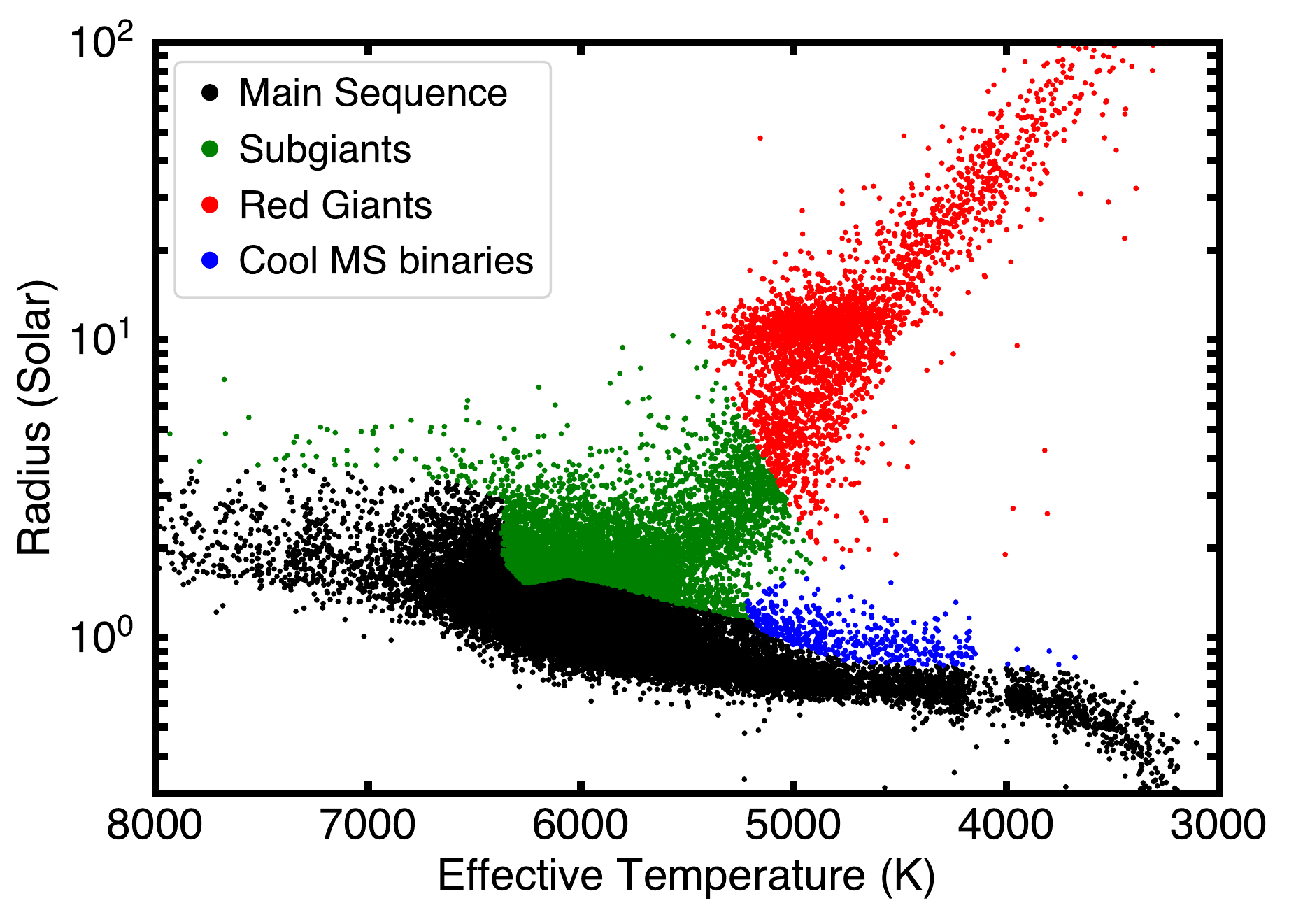}}
\caption{Evolutionary state classifications of all \kep\ targets based on physically motivated boundaries for evolutionary states (see text). We find that $\approx$\,\fdwarfs\% (\ndwarfs) of all \kep\ targets are main-sequence stars (black), $\approx$\,\fsubg\% (\nsubg) are subgiants (green), and $\approx$\,\fgiants\% (\ngiants) are red giants (red). Approximately \nbinary\ cool main-sequence stars are affected by binarity (blue).} \label{fig:class}
\end{figure}

Since the initial \kep\ target selection \citep{batalha10}, there has been growing evidence that the number of subgiants in the \kep\ Input Catalog \citep[KIC,][]{brown11} and subsequent KSPC revisions \citep{huber14,Mathur2017} have been significantly underestimated due to Malmquist bias \citep{gaidos13} and the insensitivity of broadband photometry to determine surface gravities. For example, \citet{verner11} show that radii in the KIC are underestimated by up to 50\% for a sample of subgiants with asteroseismic detections. \citet{everett13} used medium resolution spectroscopy to arrive at a similar conclusion for faint \kep\ exoplanet host stars, while surface gravities derived from granulation noise (``flicker'') suggested that nearly 50\% of all bright exoplanet host stars are subgiants \citep{bastien14}.

The revised radii using \gaia\ DR2 parallaxes presented in this work allow the first definite classification of the evolutionary states of nearly all \kep\ targets. To do this, we used solar-metallicity Parsec evolutionary tracks \citep{bressan12} to define the terminal age main sequence and base of the red-giant branch (RGB) in the temperature-radius plane, as shown in Figure \ref{fig:class}. Assuming solar metallicity means that the classifications will be only statistically accurate, but spectroscopic surveys of the \kep\ field such as the Large Sky Area Multi-Object Fiber Spectroscopic Telescope \citep[LAMOST;][]{decat15} have confirmed that the average metallicity of \kep\ targets is solar \citep{dong14}. 

To classify cool main sequence stars affected by binarity, we combined a 15\,Gyr isochrone at \feh\,=\,0.5\,dex (for warmer stars) with an empirical cut-off determined from a fiducial main sequence (for cooler stars). The latter was determined by fitting Gaussians to radius distributions in fixed \teff\ bins and fitting a fourth order polynomial to the centroid values, yielding:
\begin{equation}
\begin{split}
\log{L} = -0.69772909 + 2.1574491 x + 1.9520690 x^{2} + \\ 
16.041470 x^{3} -37.341466 x^{4}
\end{split}
\end{equation}
where $x = T_{\rm eff} /4633.78 - 1$. Based on the observed bi-modality at a given temperature we choose a cut-off of $1.4$\,$L$ to define candidate cool main-sequence binaries (blue points in Figure \ref{fig:class}). Based on the classifications shown in Figure \ref{fig:class}, we find that $\approx$\,67\% (\ndwarfs) of all \kep\ targets are main-sequence stars, $\approx$\,21\% (\nsubg) are subgiants, and $\approx$\,12\% (\ngiants) are red giants. Approximately \nbinary\ \kep\ targets are cool main-sequence binary candidates (blue). Restricting the sample to  \teff\,=\,5100--6300\,K yields a subgiant fraction of $\approx$\,31\%, and we confirmed that this fraction is relatively insensitive to apparent magnitude. While this confirms that a substantial fraction of \kep\ stars are more evolved than previously thought (see also Figure \ref{fig:radcomp}), it also demonstrates that some earlier estimates of subgiant contamination rates in the KIC and KSPC were too high, and that \kep\ did mostly target main-sequence stars. Indeed, the subgiant fractions stated above are upper limits since some stars will be affected by binarity similar to the cool main-sequence stars. The new classifications provided here will provide valuable input for planet occurrence studies, which rely on accurate stellar parameters of the parent sample \citep[e.g.][]{burke15}.

\section{Revised Properties of \kep\ Exoplanets} \label{sec:exoplanets}

\begin{figure}
\centering
\resizebox{\hsize}{!}{\includegraphics{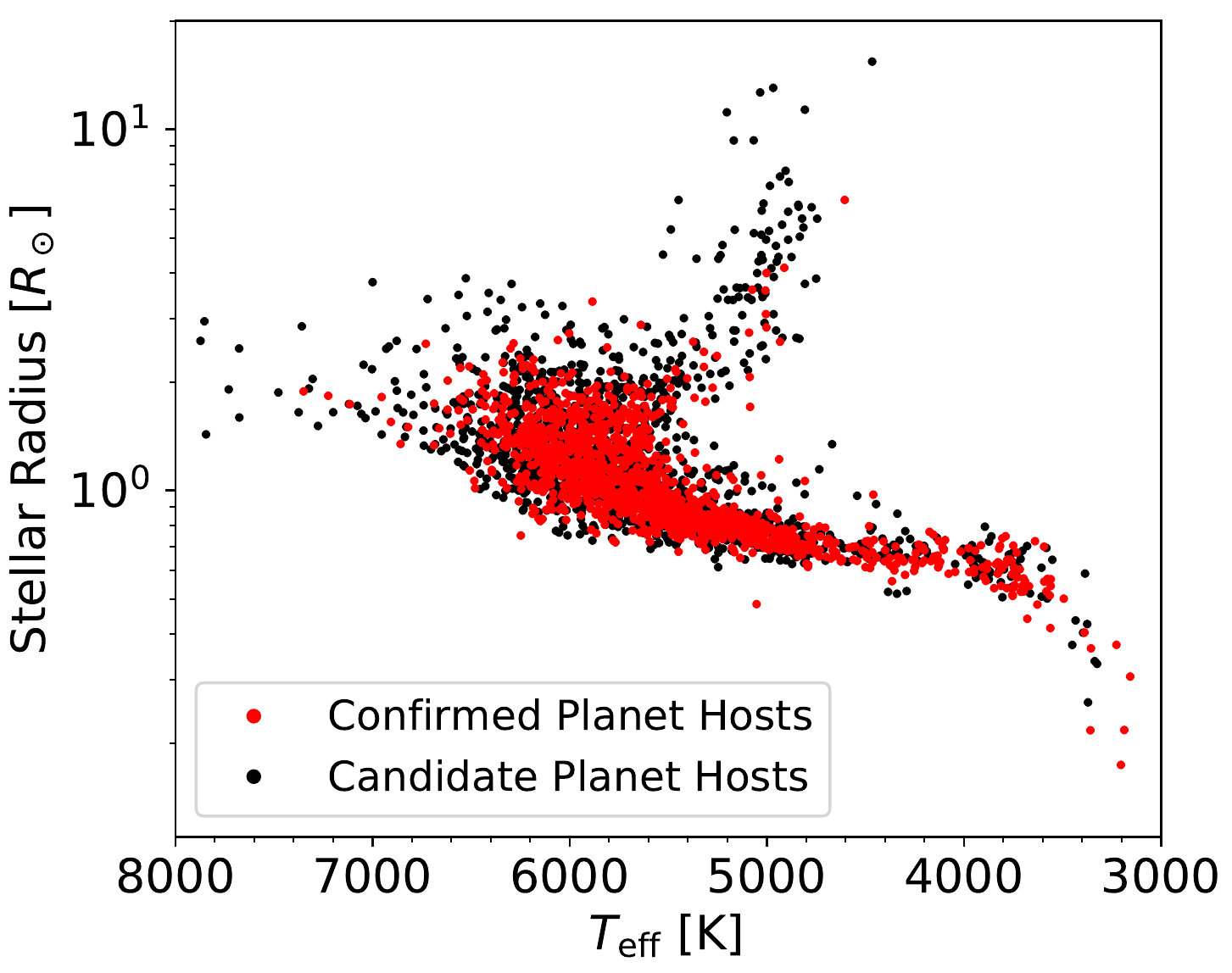}}
\caption{Hertzprung-Russell diagram displaying 1470 \kep\ confirmed planet hosts (in red) and 1524 \kep\ candidate planet host stars (in black).} \label{fig:HRHosts}
\end{figure}

\subsection{The $Gaia$ H-R Diagram of \kep\ Planet Host Stars}

Figure \ref{fig:HRHosts} displays the stellar radii and \teff\ distribution of \kep\ planet host stars, which mostly tracks the overall \kep\ population in Figure \ref{fig:HR}. While there are a similar number of confirmed (1470, red) and candidate (1524, black) planet hosts, a larger proportion of the hosts stars are more evolved. This is consistent with the expected larger number of false-positives around more evolved stars, which display larger correlated noise due to granulation \citep{sliski14,barclay14}. Several confirmed and candidate host stars fall below the main sequence and may be metal-poor subdwarfs.

\subsection{Comparison to Previous Planet Radii} \label{sec:pradcomp}

\begin{figure}
\resizebox{\hsize}{!}{\includegraphics{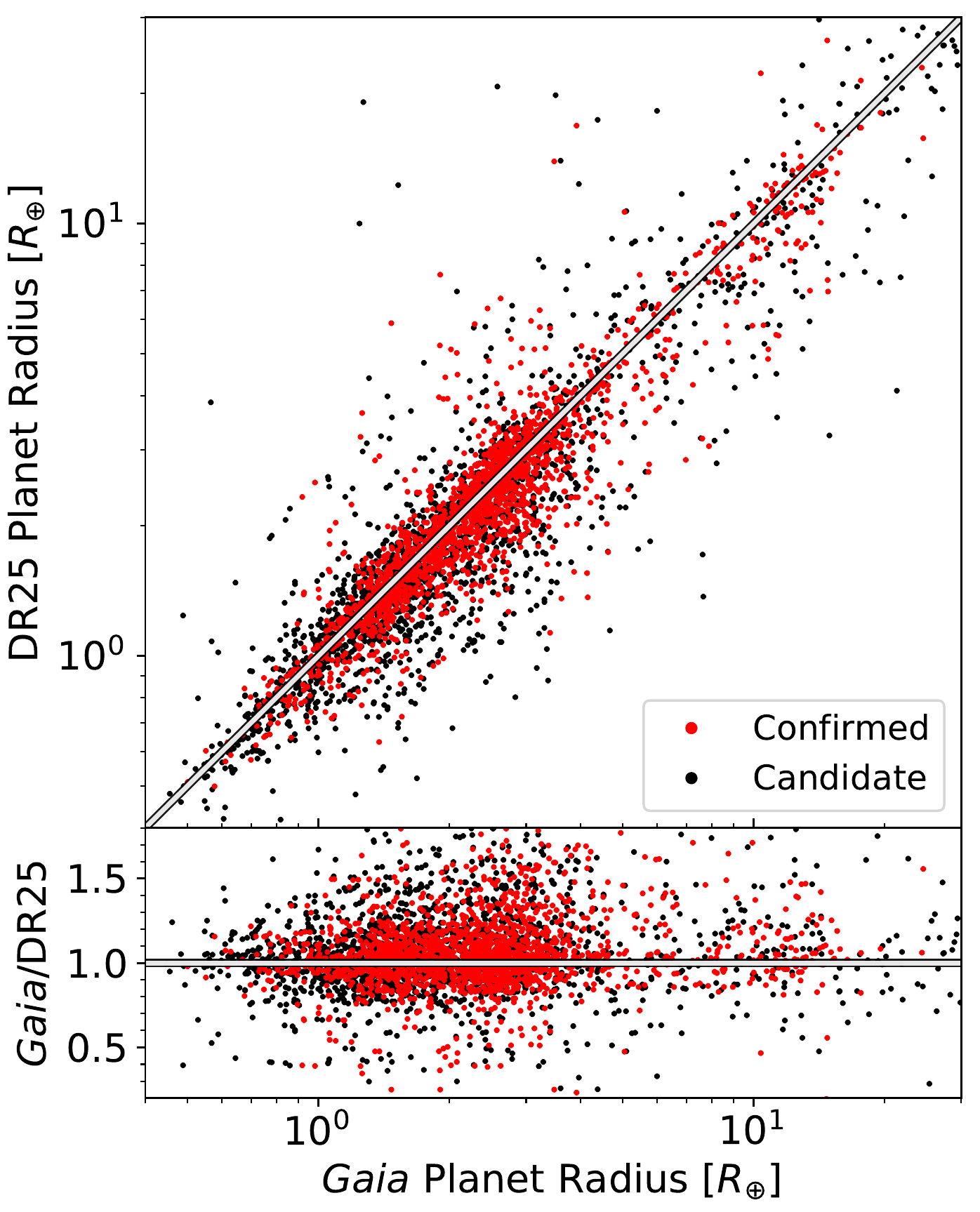}}
\caption{Planet radii calculated from stellar radii derived in this work compared to those based on stellar radii in the \kep\ DR25 Stellar Properties Catalog \citep{Mathur2017}. The red points are confirmed planets, while the black points are planet candidates. The white and black line is the 1:1 comparison between DR25 planet radii and our derived planet radii. The bottom panel shows the ratio between DR25 radii and our radii.} \label{fig:pradcomp}
\end{figure}

From the stellar radii derived above, we computed updated planet radii by utilizing the planet-star radius ratio reported in the cumulative \kep\ Object of Interest (KOI) table of the NASA Exoplanet Archive \citep{akeson13, Thompson2018} and then multiplying this ratio by our computed stellar radius. Our revised planet radii and uncertainties are given in Table \ref{tab:planets} along with a binary flag for stars with detected companions \citep[binary flag = 1,][]{Ziegler2018}. All of our data products (Tables \ref{tab:stars} and \ref{tab:planets} and additional parameters) are available at the Mikulski Archive for Space Telescopes (MAST) via \dataset[doi:10.17909/t9-bspb-b780]{http://dx.doi.org/10.17909/t9-bspb-b780}\footnote{\url{https://archive.stsci.edu/prepds/kg-radii/}}. In an attempt to quantify how much the corrections to stellar radii affect planet radii, we compare planet radii calculated using the stellar radii in KSPC DR25 and in this work in Figure \ref{fig:pradcomp}. We can see from the top panel that some planets radii change significantly with the stellar radius corrections initiated by \gaia\ DR2. The bottom panel reveals a slight systematic offset from 1--5\,\rearth, with our revised planet radii being larger. We expect this discrepancy arises because most of these planets orbit subgiant stars that were previously misclassified as dwarfs.

\begin{figure}
\resizebox{\hsize}{!}{\includegraphics{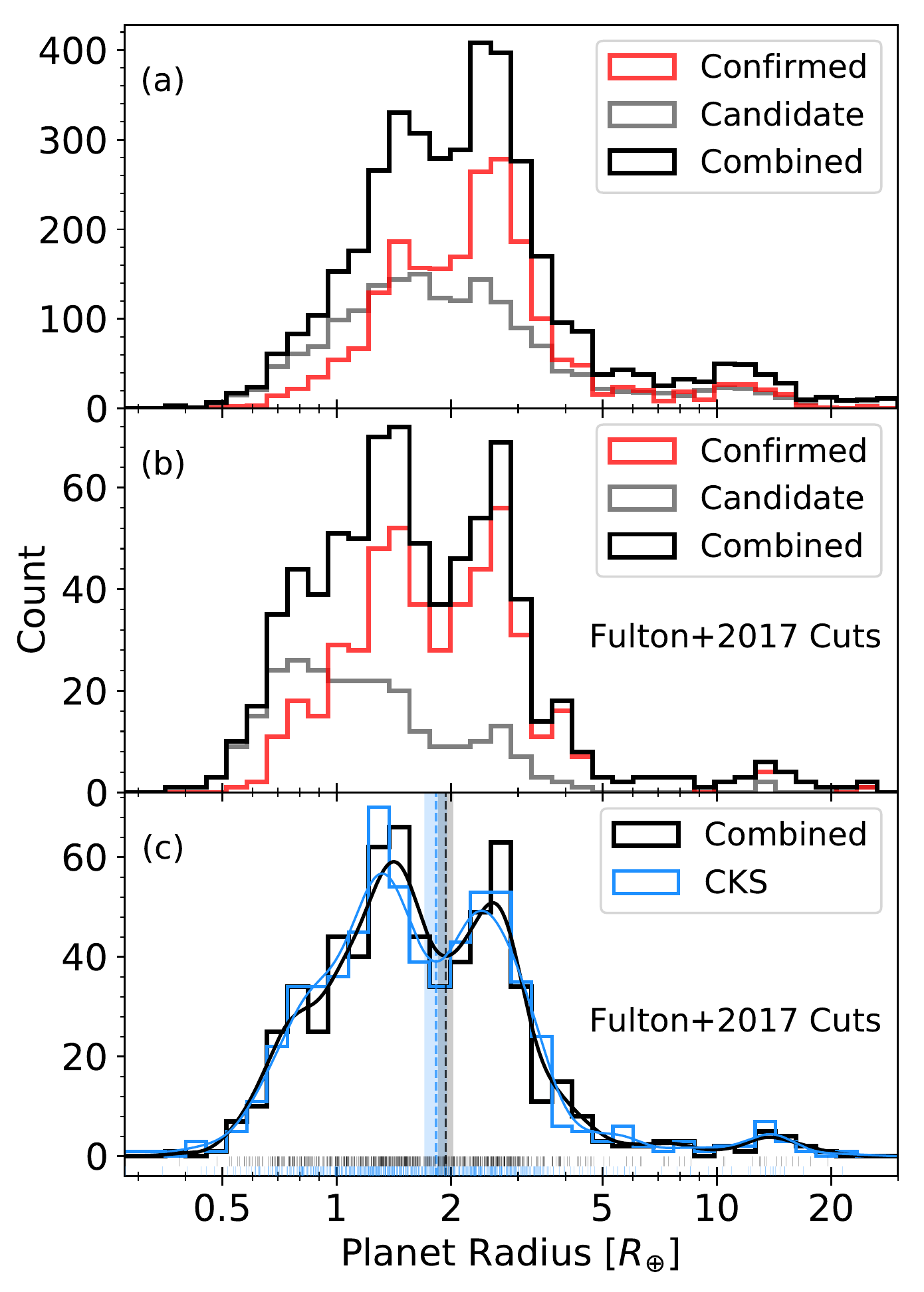}}
\caption{Distribution of \kep\ exoplanet radii computed in this work. \textit{Panel (a):} The red, gray, and black histograms contain the confirmed (2120 planets), candidate (1839 planets), and combined samples of 3959 \kep\ planets, respectively. \textit{Panel (b):} Same as Panel (a) but after performing the sample cuts described in \cite{Fulton2017}. \textit{Panel (c):} Same as Panel (b) but using only stars in the CKS sample and overplotting the CKS-derived radii in blue. Smooth lines show kernel density estimate (KDE) distributions, normalized to the total number of planets. The gap locations derived from the KDE distributions (uncorrected for occurrence rates) are 1.94\,$\pm$\,0.09\,$R_\oplus$ (this work) and 1.83\,$\pm$\,0.13\,\rearth\ (CKS).}
\label{fig:prad}
\end{figure}

In Figure \ref{fig:prad} we plot histograms of planet radii, separating candidate (gray) from confirmed (red) planets. Figure \ref{fig:prad}a includes the entire sample of 3959 planets with computed $R_{\mathrm{p}}$\,$<$\,30\,\rearth. Even from this (likely contaminated) sample, we readily recover the previously-reported gap in the radius distribution at $\sim$\,2\,\rearth\ \citep{Lopez2013,owen13}. Utilizing the precise radii of the California-\kep\ Survey \citep[CKS,][]{Petigura2017,Johnson2017}, \cite{Fulton2017} confirmed a a dearth of planets with radii $\approx$1.8\,\rearth. In addition, \cite{vaneylen2017} used asteroseismic radii to investigate the distribution of sizes of smaller planets and found a similar feature. Interestingly, our gap appears to occur at slightly larger planet radii as compared to \cite{Fulton2017}, and that the intrinsic width of the gap is not visibly increased by the more precise planet radii made possible by \gaia\ DR2 (i.e., that the width of the gap is not primarily controlled by measurement error).

Next, we implemented the same filters as in \cite{Fulton2017} to ensure a complete, well-defined population of parent stars and planets. Figure \ref{fig:prad}b, which includes 503 confirmed and 260 candidate planets, shows our ``clean'' sample after making the cuts of \cite{Fulton2017}: $K_{\mathrm{p}}$\,$<$\,14.2\,mag, 4700\,$<$\,\teff\,$<$\,6500\,K, $b$\,$<$\,0.7, $P$\,$<$\,100\,days, remove all giants and subgiants, and ignore all planets with current dispositions as false positives according to the NASA Exoplanet Archive. We see a significantly deeper gap in the confirmed sample compared to the candidate sample, and it appears to occur at the same location as the combined sample displayed in Figure \ref{fig:prad}a. Figure \ref{fig:prad}b also shows a number of very small candidate planets ($R_{\mathrm{p}}$\,$<$\,1.0\,\rearth), although we expect at least some of these planet candidates will be flagged as false positives in the future due to their low signal-to-noise ratio transits.

\begin{figure*}
\resizebox{\hsize}{!}{\includegraphics{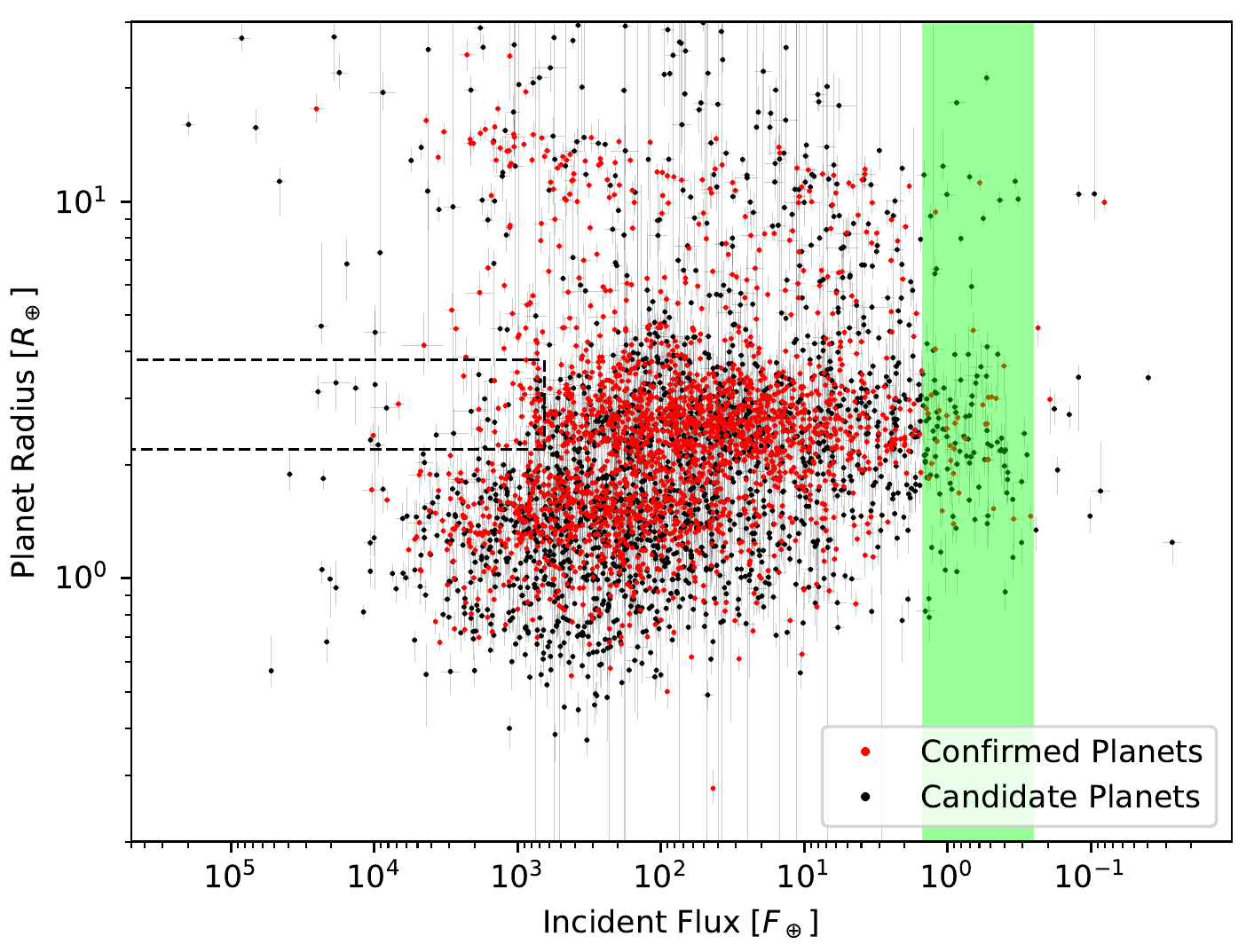}}
\caption{Planet radius versus incident flux for \kep\ exoplanets. Red and black dots are confirmed and candidate exoplanets, respectively. We also plot our asymmetric error bars in transparent gray. The dashed line box represents the extension of the super-Earth desert identified in \cite{Lundkvist2016}, while the green bar indicates the approximate optimistic habitable zone for FGK stars as detailed in \cite{Kane2016}.} 
\label{fig:insol}
\end{figure*}

Figure \ref{fig:prad}c provides a comparison of planet radii for the CKS sample of planets. The black histogram represents the planet radii computed in this work, while the blue histogram comprises those computed by the CKS team \citep{Petigura2017,Johnson2017}, both after applying the \cite{Fulton2017} filters. There are 641 planets in each histogram. We also plot a kernel density estimate (KDE) normalized to the total number of planets within each histogram. We use a Gaussian kernel for our KDEs. Below the curves, the vertical, colored ticks are the exact planet radius values that produce the color-matched curves. Finally, the dashed, colored vertical lines and the shaded regions indicate the gap location and the uncertainties, respectively, for each matched KDE.

We calculated the gap location and uncertainties by randomly drawing a planet radius value from a Gaussian distribution with a mean of its actual value and a standard deviation corresponding to its uncertainty. We then produced a KDE out of the simulated planet radii, from which we could identify the gap by finding the relative minimum between the two peaks in the simulated KDE. We repeated this process 100 times and then computed the standard deviation of the distribution of gap locations. We find the location of the gap in our distribution to be at 1.94\,$\pm$\,0.09\,$R_\oplus$, compared to 1.83\,$\pm$\,0.13\,\rearth\ for the CKS radii, where both distributions are uncorrected for occurrence rates. We thus find that the gap location derived from our radius values is slightly larger, but consistent to within 1\,$\sigma$ of previously reported values.

In addition, we quantified the effect that occurrence corrections have on the location of the gap. We did this by multiplying both of the KDEs, from 1--3\,\rearth, by a linear function so that the relative heights of the ``corrected'' KDEs match those of the super-Earth and sub-Neptune peaks in Figure 7 of \cite{Fulton2017}. The resulting changes ($\approx$\,--0.07\,\rearth\ or smaller) shift both gap locations to smaller values, but both are within our reported uncertainties.

\subsection{Distributions of Planets with Radius and Stellar Irradiation} \label{sec:distributions}

Figure \ref{fig:insol} plots planet radii versus orbit-averaged incident stellar irradiation $F$ in Earth units, using the revised host star parameters and assuming the semi-major axes reported in the NASA Exoplanet Archive, and circular orbits. Planets with slight eccentricities, or near-circular orbits, do not experience a large difference in their incident flux compared to planets on perfectly circular orbits since $F$\,$\propto$\,$1/\sqrt{1-e^2}$ \citep{Mendez2017}. We do not account for possible differences in host star mass derived from pre-\gaia\ DR2 stellar radius values and those reported here, as those effects will be much smaller than the change in luminosity and would require isochrone fitting. Several features in this diagram that have been previously described in the literature become more distinct with the improved precision in stellar and planet properties enabled by \gaia.

\begin{figure*}
\resizebox{\hsize}{!}{\includegraphics{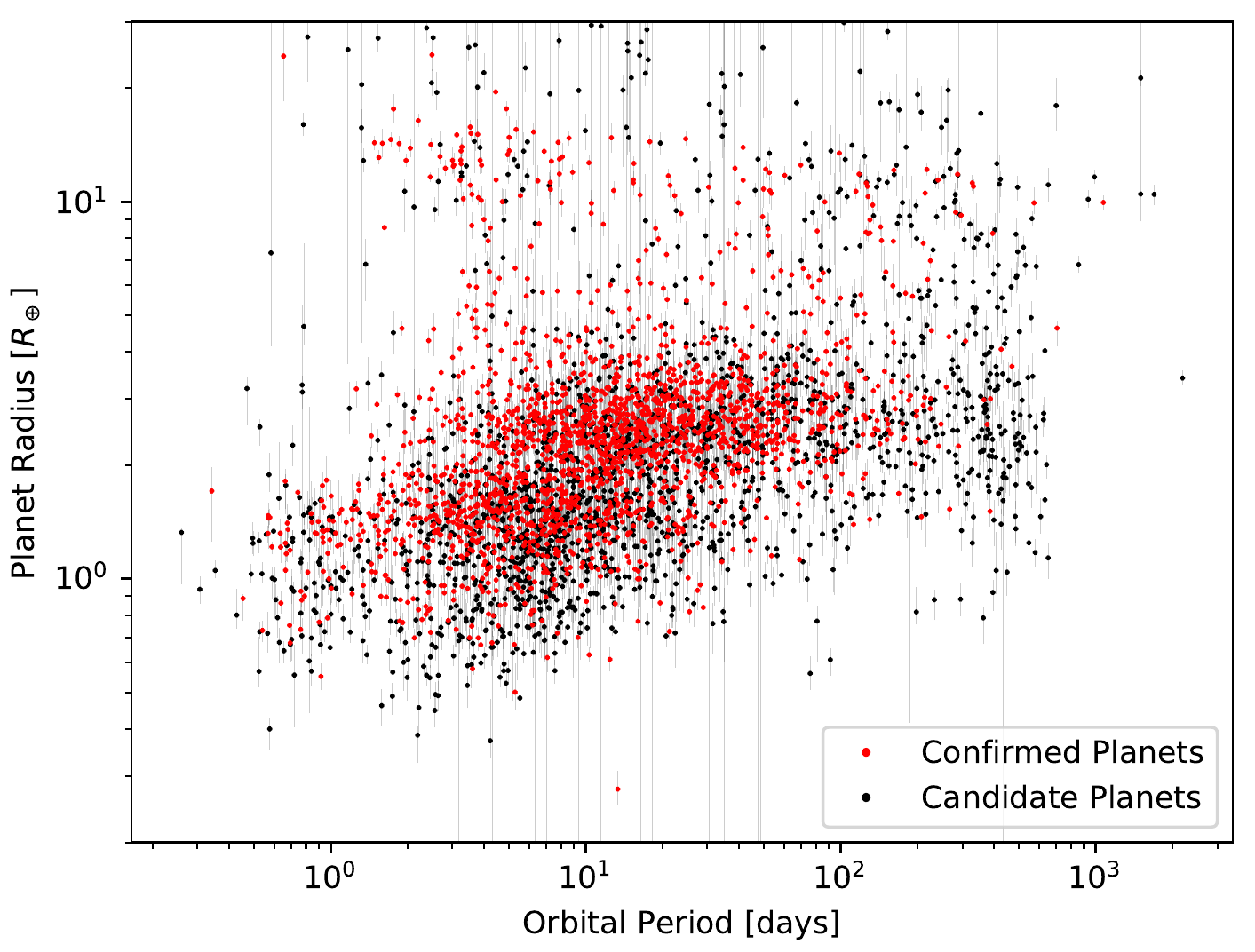}}
\caption{Same as Figure \ref{fig:insol}, but with orbital period in place of incident flux as the x-axis.} 
\label{fig:rad-period}
\end{figure*}

\subsubsection{The Small Planet Radius Gap} \label{sec:gap}

As shown in Figures \ref{fig:prad} and \ref{fig:insol}, our revised parameters confirm the bimodal distribution of planet radii, with a gap or ``evaporation valley" between the two peaks. The depth of the gap depends on stellar irradiance, with a clear gap just above 2\,\rearth\ for $F$\,$>$\,200\,\fearth, the absence of an obvious gap at 30--200\,\fearth, and a less distinct deficit of planets in this size range at $F$\,$<$\,30\,\fearth. Similar to the integrated value reported in Section \ref{sec:pradcomp}, the gap in the high-irradiance regime appears at slightly larger planet radii than in \cite{Fulton2017} (see their Figure 8). We suspect that this difference arises from sample selection and systematically smaller CKS stellar radii compared to Gaia radii for slightly evolved stars, as we find a $\sim$\,5\% systematic underestimation in CKS stellar radii \citep[see also Figure 3 in][]{Fulton2018}.

The gap is predicted by models in which photoevaporation due to X-ray and ultraviolet (XUV) radiation, more common early in a star's lifetime, removes the light molecular weight envelopes of planets. The relationship between planet mass, surface gravity, and loss rate means that the envelopes of intermediate-size planets are efficiently stripped, producing distinct populations of rocky planets and more massive planets that retain their envelopes \citep{Owen2017,Jin2018}. This process is more efficient at high irradiance, which explains the prominence of the gap in that regime. Also according to models, the location of the gap constrains the composition of the residual planet ``cores.'' A gap at a larger radius would mean a greater contribution by lower density ices. For example, \cite{Jin2018} finds that an ``evaporation valley'' at 1.6\,\rearth\ corresponds to an Earth-like composition of silicates and metals, so a valley at a larger radius implies a significant ice component.

Recent investigations have revealed that the location of the radius gap depends on host star mass \citep{Fulton2018,Wu2018}. \cite{Fulton2018} also investigated the distribution of planets in radius-orbital period space, and did not find a strong dependence of the orbital period distribution on stellar mass (and thus main-sequence luminosity). This supports XUV-driven photoevaporation as the dominant mechanism sculpting the radii of the exoplanet population, while other mechanisms such as core-powered mass loss \citep{Ginzburg2016,Ginzburg2018} seem less important. Deriving stellar masses for the entire Kepler sample will be left for future work, but we note that some differences in the radius distributions may be due to the fact that our sample includes host stars spanning all spectral types (including M dwarfs).

\subsubsection{Hot Super-Earth ``Desert"} \label{sec:des}

Our revised radius and irradiance values confirm the existence of a deficit or ``desert" of super-Earth to Neptune-size planets at high irradiance \citep{owen13}, i.e.\,with 2.2\,$<$\,$R_{\mathrm{p}}$\,$<$\,3.8\,\rearth\ and $F$\,$>$\,650\,\fearth\ \citep{Lundkvist2016}. This desert could be a consequence of photoevaporation of the hydrogen-helium envelopes of sub-Neptune-size planets at stellar irradiance levels more extreme than that which produced the gap \citep{owen16,Lehmer2017}, but \citet{ionov18} suggests that some other mechanism must be present. Alternatively, the desert could be explained if only rocky planets, not mini-Neptunes, form close to stars because the inner disk is depleted in gas and volatiles \citep{lopez16}. For these two mechanisms, the underlying important variable is the irradiation by the host star and the orbital period/semi-major axis, respectively. These variables are weakly related at the population level because of the wide range of luminosities (five orders of magnitude) of the host stars in the \kep\ sample. In a plot of radius versus \emph{orbital period} (Figure \ref{fig:rad-period}) the boundaries of the desert are also apparent. However, the transition to the desert at short orbital periods for sub-Neptunes is not as abrupt compared to the marginally sharper drop-off in planets at $F$\,$>$\,650\,\fearth\ in Figure \ref{fig:insol}, indicating that orbital period is not the underlying ``master" variable.

Additionally, we find that the ``hot desert'' \citep{Lundkvist2016} is not so empty after all. Forty-six confirmed and 28 candidate planets fall within this range. About half are close to the 650\,\fearth\ boundary, and our refined parameters suggest that a distinct edge exists at $\approx$\,$10^3$\,\fearth, but 13 confirmed and two candidate planets are more than 2\,$\sigma$ interior to all the edges of the desert. The host stars of these desert dwellers are almost exclusively subgiant stars more massive than the Sun that are evolving towards or at the red giant branch. This is in contrast with the smaller planets in this irradiance range, which orbit both evolved and main sequence stars, and larger (sub-Jovian and Jovian) hot planets, which are found around subgiants with a range of masses. Transit detection bias can explain the large numbers of smaller hot planets around dwarf stars, but not the absence of mini-Neptunes. If the hot mini-Neptunes were the transient remnant of a depleted population we would expect their host stars to be younger than average, but their evolutionary state suggests that they are older.

\citet{Lopez2017} finds that the absence of sub-Neptunes in the ``desert'' can be explained if planets of this size have hydrogen-helium envelopes, but \emph{not} substantial envelopes of high molecular weight volatiles (e.g. H$_2$O) which would be retained. The exceptions here suggest that at least some of these objects do have high molecular weight envelopes, and/or that they have evolved from a different planet population. One explanation for these interlopers is that they are the product of evaporation of still larger objects, i.e.\ sub-Jovian or even Jupiter-size planets that have lost much of their envelopes. \citet{dong18} find that the metallicities of host stars of hot Neptunes are distributed similarly to that of the host stars of hot Jupiters, suggesting a relationship between the two populations. One long-standing idea is that hot Neptunes are the product of massive photoevaporation of a giant planet's envelope \citep{baraffe05}.

Another potential explanation for the presence of planets within the ``desert'' is guided by the theory discussed in \cite{Owen2017}. Because the hosts of these desert-dwelling planets are probably more massive, which is why they have subsequently evolved into subgiants, the integrated XUV radiation from the main sequence progenitors was lower due to the shorter-main sequence lifetime and inefficient dynamo operation in star without a convective-radiative boundary ($M$\,$>$\,1.3\,\msun). The dearth of XUV irradiation from these stars allowed their planets to retain low-molecular weight envelopes.

\subsubsection{Inflated Hot Jupiters} \label{sec:hotj}

Another feature revealed by Figure \ref{fig:insol} is the well-known trend of increasing giant planet radius with increasing stellar irradiance \citep[e.g.][]{Burrows2000,Demory2011,Laughlin2011}. Confirmed planets with inflated ($>$\,1.2\,$R_{\mathrm{J}}$) radii are numerous at $F$\,$>$\,150\,\fearth, consistent with previous work and planet inflation theory \citep{Lopez2016}. These include giant planets orbiting subgiants and low-luminosity red giants hosts, including previously discovered examples \citep{Grunblatt2016,Grunblatt2017}.  Giant planet inflation by irradiation could arise from different mechanisms of transport of heat to the planet interior, or suppression of cooling \citep{Lopez2016}. We identified four confirmed inflated giant planets at \emph{low} ($<$\,150\,\fearth) irradiation: \kep-447b, \kep-470b, \kep-706b, and \kep-950b, but of these only \kep-470b satisfy the ``cool" inflated planet at more than two sigma significance.  Despite the disposition listed in DR25, \kep-470b was identified by \citet{Santerne16} to be an eclipsing binary based on radial velocities.  

\subsubsection{Habitable Zone Planet Candidates} \label{sec:hab}

Finally, we identify candidate and confirmed planets within the circumstellar ``habitable zone" where surface temperatures on an Earth-size planet with an Earth-like composition, geology, and geochemistry would permit liquid water. Following \citet{Kane2016}, we adopt the ``optimistic" definition $0.25$\,$<$\,$F$\,$<$\,1.50\,\fearth\ and illustrate this as the green bar in Figure \ref{fig:insol}. In this habitable zone we identify 34 confirmed planets and 109 candidate planets. Of these, 30 planet candidates and 8 confirmed planets have \rp\,$<$\,2\,\rearth: \kep-62e, \kep-62f, \kep-186f, \kep-440b, \kep-441b, \kep-442b, \kep-452b \citep[but see also][]{Mullally18}, and \kep-1544b.  These candidate planets should be priority targets for follow-up observations to vet the planets and better characterize the host stars, so as to better establish the occurrence of potential Earth-like planets $\eta_{\oplus}$. 

\section{Summary and Conclusions} \label{sec:conc}

We presented a re-classification of stellar radii for \nstars\ observed by the \kep\ Mission by combining \gaia\ DR2 parallaxes with the DR25 \kep\ Stellar Properties Catalog \citep[KSPC,][]{huber14,Mathur2017}. The typical precision of stellar radii is $\sim$\,8\%, a factor of 4-5 better than previous estimates in the KSPC. Based on the revised stellar radii, we have furthermore re-derived radii for 2123 confirmed planets 1922 planet candidates discovered by \kep. Our main conclusions are as follows:

\begin{itemize}

\item We find that \fdwarfs\% (\ndwarfs) of all \kep\ targets are main-sequence stars, \fsubg\% (\nsubg) are subgiants, and \fgiants\% (\ngiants) are red giants. While many radii are revised to larger values, this demonstrates that previous findings of large subgiant contaminations in the \kep\ Input Catalog (KIC) and KSPC were likely overestimated, and that the \kep\ parent population indeed consists mostly of main-sequence stars.

\item We find evidence for binarity in \nbinary\ cool main-sequence stars ($\sim$\,2\% of the overall sample) based on their inflated radii in the H-R diagram. This demonstrates that \gaia\ parallaxes can be used to efficiently identify binary stars, and we encourage follow-up observations of the binary candidates identified in our work (see Table \ref{tab:stars}).

\item We confirm the gap in the radius distribution of small \kep\ planets \citep{Fulton2017}. Our observed gap for the \cite{Fulton2017} sample of 1.94\,$\pm$\,0.09\,$R_\oplus$ (without occurrence rate corrections, which would shift the value by $\approx$\,--0.07\,\rearth) is at a slightly larger radius but consistent to within 1\,$\sigma$ with previously reported planet radius distributions. The planet radius--incident flux plot reveals the gap over a wide range of incident fluxes, with the largest gap occurring at 200\,\fearth. The location of the gap has important implications for planet formation and evolution theory, as it can constrain planetary core compositions.

\item Planets do reside in a region of radius-irradiance space previously referred to as the ``hot super-Earth desert'' \citep{Lundkvist2016}. We identify 74 stars hosting 46 confirmed planets and 28 planet candidates that receive $>$\,650\,\fearth\ and have radii between 2.2 and 3.8\,\rearth. However, we confirm that there is a clear paucity of super-Earths in the desert regime, especially at incident fluxes $>$\,1000\,\fearth.

\item We observe a clear inflation trend for hot Jupiters, where inflated planets become numerous at an irradiation level $>$\,150\,\fearth. We identify a few confirmed planets that may be inflated Jupiters at incident fluxes $<$\,150\,\fearth\ (\kep-447b, \kep-470b, \kep-706b, and \kep-950b), but find that the most promising case (\kep-470b) was previously reported as an eclipsing binary.

\item We identify 34 confirmed planets and 109 planet candidates within the habitable zone. Of these planets, 30 planet candidates and 8 confirmed planets have \rp\,$<$\,2\,\rearth: \kep-62e, \kep-62f, \kep-186f, \kep-440b, \kep-441b, \kep-442b, \kep-452b \citep[but see also][]{Mullally18}, and \kep-1544b. These systems in particular represent a high priority sample for ground-based follow-up.

\end{itemize}

We have applied $Gaia$ DR2 measurements to \kep\ stars and their planets and identified several patterns in the distribution of both stars and planet properties that suggest avenues of future investigation. In this work, we have restricted our refinement of stellar properties to their radii and luminosities, but future work will exploit precise \gaia\ parallaxes by applying stellar evolution models to infer surface gravities, densities, masses and ages.  Planet populations are expected to evolve with time as a result of cooling and contraction of envelopes, photo-evaporation of atmospheres, and mutual dynamical scattering. It may also be possible to observe this evolution with sufficiently well-selected and characterized samples of old and young stars and planetary systems, \citep[e.g.][]{Mann2017,Berger2018}. The unprecedented parallaxes provided by $Gaia$ will continue to reveal new and interesting information about stars and their companions, and more in-depth analyses of singular systems will inevitably lead to some unpredicted discoveries.

\begin{deluxetable*}{ccccccccccccc}
\tabletypesize{\scriptsize}
\tablenum{1}
\tablewidth{0pt}
\tablecolumns{12}
\tablecaption{Revised Parameters of \kep\ Stars}
\tablehead{
\colhead{KIC ID} & \colhead{$Gaia$ DR2 ID} & \colhead{\teff\ [K]} & \colhead{$\sigma_{\teffeq}$ [K]} & \colhead{$d$ [pc]} & \colhead{$\sigma_{d+}$ [pc]} & \colhead{$\sigma_{d-}$ [pc]} & \colhead{$R_\star$ [$R_\odot$]} & \colhead{$\sigma_{R_{\star+}}$ [$R_\odot$]} & \colhead{$\sigma_{R_{\star-}}$ [$R_\odot$]} & \colhead{$A_V$ [mag]} & \colhead{Evol.\ Flag} & \colhead{Bin.\ Flag}}
\def\arraystretch{1.0}
\startdata
757076&2050233807328471424&5164&181&658.465&21.419&20.163&3.986&0.324&0.293&0.273&1&0\\
757099&2050233601176543104&5521&193&369.374&3.708&3.645&1.053&0.078&0.071&0.120&0&0\\
757137&2050230543159814656&4751&166&570.715&8.271&8.060&13.406&1.004&0.916&0.230&2&0\\
757280&2050230611879323904&6543&229&824.791&15.079&14.586&2.687&0.205&0.186&0.323&0&0\\
757450&2050231848829944320&5306&106&835.371&18.423&17.692&0.962&0.047&0.044&0.298&0&0\\
892010&2050234975566082176&4834&169&1856.534&86.437&79.285&14.826&1.302&1.178&0.258&2&0\\
892107&2050234696381511808&5086&178&941.305&20.518&19.713&4.334&0.334&0.303&0.186&2&0\\
892195&2050234735047928320&5521&193&480.822&3.850&3.800&0.983&0.073&0.066&0.141&0&0\\
892203&2050236521754360832&5945&208&555.165&4.828&4.759&1.022&0.076&0.069&0.124&0&0\\
892667&2050232329866306176&6604&231&1175.938&21.455&20.754&2.207&0.168&0.153&0.352&0&0\\
892675&2050232329866320512&6312&221&584.442&4.837&4.772&1.052&0.078&0.071&0.175&0&0\\
\enddata
\tablecomments{KIC ID, $Gaia$ DR2 ID, \teff, distance, stellar radii, extinction, evolutionary flag, and binary flag (and errors, where reported) for our sample of \nstars\ \kep\ stars. The evolutionary flags are as follows: 0 = main sequence dwarf, 1 = subgiant, and 2 = red giant. The binary flags are as follows: 0 = no indication of binary, 1 = binary candidate based on \gaia\ radius only, 2 = AO-detected binary only \citep{Ziegler2018}, and 3 = binary candidate based on \gaia\ radius and AO-detected binary. See Figure \ref{fig:class} for stars with evolutionary state flags = 0--2 (black, green, and red, respectively) and binary flags = 1 or 3 (blue). A slice of our derived parameters is provided here to illustrate the form and format. The full table, in machine-readable format, can be found online.} \label{tab:stars}
\end{deluxetable*}

\begin{deluxetable*}{ccccccccc}
\tabletypesize{\scriptsize}
\tablenum{2}
\tablewidth{0pt}
\tablecolumns{8}
\tablecaption{Revised Parameters of \kep\ Exoplanets}
\tablehead{\colhead{KIC ID} & \colhead{KOI ID} & \colhead{\rp\ [\rearth]} & \colhead{$\sigma_{\rpeq+}$ [\rearth]} & \colhead{$\sigma_{\rpeq-}$ [\rearth]} & \colhead{$F_\mathrm{p}$ [\fearth]} & \colhead{$\sigma_{F+}$ [\fearth]} & \colhead{$\sigma_{F-}$ [\fearth]} & \colhead{Binary Flag}}
\def\arraystretch{1.0}
\startdata
10797460&K00752.01&2.316&0.156&0.134&104.641&8.011&7.367&0\\
10797460&K00752.02&2.898&0.955&0.207&10.186&0.780&0.717&0\\
10854555&K00755.01&2.308&0.410&0.230&652.435&55.222&50.242&0\\
10872983&K00756.01&4.600&0.721&0.334&122.778&14.831&13.014&0\\
10872983&K00756.02&3.268&0.341&0.337&457.318&55.241&48.473&0\\
10872983&K00756.03&1.874&0.510&0.217&863.776&104.338&91.555&0\\
10910878&K00757.01&4.879&0.284&0.260&21.637&1.680&1.541&1\\
10910878&K00757.02&3.272&0.207&0.180&6.162&0.478&0.439&1\\
10910878&K00757.03&2.245&0.150&0.125&76.217&5.917&5.428&1\\
11446443&K00001.01&14.186&0.602&0.570&903.864&20.395&20.041&1\\
\enddata
\tablecomments{KIC ID, KOI ID, planetary radii, incident fluxes (and errors where reported), and AO-detected companion flags \citep{Ziegler2018} of our sample of 4045 \kep\ confirmed/candidate planets. A slice of our derived parameters is provided here to illustrate the form and format. The full table, in machine-readable format, can be found online.} \label{tab:planets}
\end{deluxetable*}

\acknowledgments 

We gratefully acknowledge everyone involved in the \gaia\ and \kep\ missions for their tireless efforts which have made this paper possible. We also thank the reviewer for helpful feedback which improved this paper, and Erik Petigura and BJ Fulton for helpful discussions. T.A.B. and D.H. thank Savita Mathur for providing supplementary material for the DR25 stellar properties catalog. T.A.B. and D.H. acknowledge support by the National Science Foundation (AST-1717000) and the National Aeronautics and Space Administration under Grants NNX14AB92G issued through the \kep\ Participating Scientist Program. This work has made use of data from the European Space Agency (ESA) mission {\it Gaia} (\url{https://www.cosmos.esa.int/gaia}), processed by the {\it Gaia} Data Processing and Analysis Consortium (DPAC, \url{https://www.cosmos.esa.int/web/gaia/dpac/consortium}). Funding for the DPAC has been provided by national institutions, in particular the institutions participating in the {\it Gaia} Multilateral Agreement. This publication makes use of data products from the Two Micron All Sky Survey, which is a joint project of the University of Massachusetts and the Infrared Processing and Analysis Center/California Institute of Technology, funded by the National Aeronautics and Space Administration and the National Science Foundation. This research has made use of NASA's Astrophysics Data System. This research was made possible through the use of the AAVSO Photometric All-Sky Survey (APASS), funded by the Robert Martin Ayers Sciences Fund. This research made use of the cross-match service provided by CDS, Strasbourg. This research has made use of the NASA Exoplanet Archive, which is operated by the California Institute of Technology, under contract with the National Aeronautics and Space Administration under the Exoplanet Exploration Program.

\vspace{5mm}

\software{\texttt{astropy} \citep{astropy},
		  \texttt{dustmaps} \citep{Green2018},
		  \texttt{GNU Parallel} \citep{Tange2018},
		  \texttt{isoclassify} \citep{huber17}, 
		  \texttt{Matplotlib} \citep{Matplotlib},
          \texttt{mwdust} \citep{bovy16}, 
          \texttt{Pandas} \citep{Pandas}, 
          \texttt{SciPy} \citep{Scipy}}

\bibliography{ApJDR2Radii}

\end{document}